\documentclass[runningheads]{llncs}

\def\doi#1{\href{https://doi.org/\detokenize{#1}}{\url{https://doi.org/\detokenize{#1}}}}
\usepackage{graphicx}
\usepackage[utf8]{inputenc}
\usepackage[T1]{fontenc}
\usepackage{cite}
\usepackage{amsmath,amssymb,amsfonts}
\usepackage{algorithmic}
\usepackage{textcomp}
\usepackage{enumitem}
\usepackage{booktabs}
\usepackage[table, svgnames]{xcolor}
\usepackage{multirow, makecell}
\usepackage{hyperref}
\usepackage{tablefootnote}
\usepackage{listings}
\usepackage{chemformula}

\begin{document}
\title{Application-Oriented Selection of Privacy Enhancing Technologies\thanks{The final publication is available at link.springer.com (\url{https://doi.org/10.1007/978-3-031-07315-1_5})}}

\author{Immanuel Kunz \and
Andreas Binder}
\authorrunning{I. Kunz and A. Binder}
\institute{Fraunhofer AISEC \\
\email{\{firstname.lastname\}@aisec.fraunhofer.de}}

% \author{First Author \and
% Second Author \and
% Third Author}
% \institute{Anonymous Institute \\
% \email{email@institute.com}}
%
\maketitle              % typeset the header of the contribution
\begin{abstract}
% \item Research in privacy engineering has proposed various contributions in requirements engineering, threat modeling, and selection and classification of PETs. 
% Various approaches have been proposed to simplify the fulfillment of privacy requirements in software architectures, e.g. design patterns and threat modeling approaches.
% Privacy engineering comprises a number of activities 
To create privacy-friendly software designs, architects need comprehensive knowledge of existing privacy-enhancing technologies (PETs) and their properties. 
Existing works that systemize PETs, however, are outdated or focus on comparison criteria rather than providing guidance for their practical selection. % rather than classifying PETs with the goal of developing application-oriented selection guidance. % focus on finding common properties
% focus on single criteria, like the targeted privacy goal, or describe PET properties in a way that is not guiding their selection. 
% In existing works, however, the classification of PETs is usually approached rather theoretically, with a focus on characteristics that generic technologies share. This approach does not necessarily simplify their selection. Rather, an application-oriented classification is required.
In this short paper we present an enhanced classification of PETs that is more application-oriented than previous proposals. It integrates existing criteria like the privacy protection goal, and also considers practical criteria like the functional context, a technology's maturity, and its impact on various non-functional requirements.
% \item We also identify performance as a crucial factor for PET selection and outline steps for future work to develop a framework for the comparative evaluation of PETs regarding their performance.
We expect that our classification simpliefies the selection of PETs for experts and non-experts. % We hope that our classification will simplify the selection of PETs also for non-experts and provides the basis for automated support tools.
\end{abstract}

\section{Introduction}
\label{introduction}
%!TEX root = ../selecting-pets.tex

% \item Privacy engineering is gaining importance, but its application in day-to-day software engineering is not trivial.
A decisive activity in privacy engineering is the selection of appropriate Privacy Enhancing Technologies (PETs), for example to fulfill requirements or mitigate risks in goal-based or risk-based engineering methods respectively. While this step is highly application-specific, it can be approached systematically, since common decision criteria for PET-selection exist.
% \item After requirements engineering, a crucial step is to select suitable Privacy Enhancing Technologies (PETs) for the achievement of goals, or the mitigation of risks. While this step is highly application-specific, it can be approached systematically, since the decision criteria for PET selection are known.
For example, one criterion that can guide engineers in their design decisions is the privacy goal that is targeted by a PET, such as anonymity or undetectability. % concrete example: k-anonymity primarily targets linkability; differential privacy targets non-repudiation, but also linkability

Existing works have proposed different systematizations of PETs in the past. The LINDDUN methodology~\cite{deng2011privacy}, for instance, categorizes PETs using their privacy protection goal, and differentiates between PETs that target transactional and contextual data~\cite{linddunwebsite}. Heurix et al.~\cite{heurix2015taxonomy} categorize PETs, e.g., regarding the trust scenario they target and their involvement of a trusted third party.
Yet, these systematizations do not sufficiently take into account the practical context in which PETs are selected: they often omit the PET's functional context it can be applied in, as well as other practical criteria. Also, they are partly outdated.

In this short paper, we develop a new PET classification that is more application-oriented. Our classification builds upon previous proposals, integrating some of their criteria~\cite{deng2011privacy,heurix2015taxonomy,al2021land} like the technology's targeted privacy protection goal and its impact on other non-functional requirements. It furthermore includes the PET's functional context, as well as prioritization criteria, like the maturity of the technology. 
% We propose a number possible functional contexts. % which we derive from an analysis of cloud services. % which we expect to be representative...
% TODO adjust number
We present a classification of 30 PETs that we have done according to the proposed criteria.

% TODO: what are we evaluating/demonstrating/showing exactly? Usefulness/effectiveness/applicability/...?
To demonstrate the effectiveness of our classification, we compare it to the one proposed by LINDDUN based on a use case.
We expect that our work can support engineers in selecting appropriate PETs, and motivate PET developers to evaluate their technologies according to our criteria, making them more comparable and usable for architects and engineers.

% In summary, we present the following contributions:
% \begin{itemize}
% 	\item A list of application-oriented criteria for PET-selection,
% 	\item A classification of existing PETs following our criteria.
% \end{itemize}

\section{Classification}
\label{classification}

Various approaches to selecting appropriate PETs have been proposed in the past, but the adequate application of PETs can be even more challenging than their selection, for example due to the development effort involved in applying a PET in a certain environment. 
% ~\cite{yskout2015security}
Our goal is therefore to develop an application-centered classification of PETs that anticipates the challenges that a PET can cause when it is applied. In the following, we describe and motivate the criteria we have included in our classification. 

\subsection{Motivating Example}
To motivate the choice of criteria, we lay out the following scenario of an engineer developing an architecture for a generic privacy-friendly cloud service, assuming that it is representative for the decisions architects and developers have to make when implementing privacy requirements.
This generic service allows users to register, authenticate, provide personal data to the service (e.g. names, addresses), interact with other users of the service (e.g. sending messages), and to retrieve data that is stored in the service's storages. 

An engineer creates a threat model for an initial design of the service which reveals a detectability threat for the service's users: since their messages to other users are observable by the service and possibly external attackers, their relationships can be identified. A requirement is therefore elicited which states that users' messages should be undetectable. % (to some extent).
% computing: HE vs enclaves vs MPC % messaging: detectability -> VPN (TTP), filter+auth, P2P, steganography, random dummy traffic
A PET that mitigates this threat has to match several criteria: % to be a good fit for this scenario: 

First, it has to fit the functional context of the service. This includes the targeted privacy protection goal, such as anonymity or undetectability. To ensure suitability for the scenario, however, the targeted privacy protection goal is not sufficient; also, the PET has to fit the functional requirements of the application. In a messaging scenario, for example, different PETs are applicable than in an information retrieval or computing scenario.
Ideally, the PET should furthermore not only target the correct privacy goal, but should also be measurable in the achieved privacy gain. This way, a more comparable and reproducible selection is facilitated.
	
Second, various non-functional properties can play a role. For example, the PET should be mature enough to be applied and maintained by the development team: applying a certain technology can otherwise imply large efforts for configuration and further development of the technology. 
To be able to weigh different usable PETs against each other, also further properties are of interest: PETs can have an impact on the overall architecture, on the utility of transactional data, and the performance of the respective interaction.
% For example, the PET has to be implementable by the development team, i.e. it should not require more expertise than is available in the team. Second, it should not incur too much cost, e.g. in its set-up or its continuous maintenance, or introduce new risks/threats. Furthermore, the overhead a PET introduces is crucial to evaluate its applicability.
% These criteria may not only be used to filter certain PETs, but to prioritize a set of applicable ones according to, e.g., cost factors.
	
% Third, its privacy gain has to be measurable, so the quality of the solution can be quantified and monitored over time.

% \begin{itemize}
% 	\item When an engineer is faced with a set of privacy requirements, e.g. make a service interaction undetectable, the following questions have to be answered:
% 	\begin{itemize}
% 		\item Which PETs target undetectability?
% 		\item Which PETs fit the functional context of application?
		% \item Are there ready-to-use implementations available or are adaptations and
		% \item Will the PET incur a time effort for maintenance?
	% 	\item May the PET have an impact on the architecture design or does it only require an update to an existing component?
	% 	\item How mature is the technology? Which implementations exist and do they have to be adapted?
	% \end{itemize}
	% \item Only when these questions have been answered, a PET that is appropriate for the application at hand can be selected. 
% \end{itemize}

In the following we describe the criteria we use to cover these considerations, and classify a list of PETs accordingly.

% TODO further criteria: type of data (in motion/at rest) (consider, however, that sometimes data can be buffered to create an ``at rest'' scenario out of data in motion) % TODO source
% TODO Further NFRs: modifiability, extensibility, testability, scalability (O(n)), reusability, ``performance impact'' --> how to justify the selection?

\subsection{Criteria}
\subsubsection{Privacy Protection Goal}
One of the most meaningful criteria for selection is the targeted privacy protection goal. To that end we use the goals proposed by the LINDDUN methodology~\cite{deng2011privacy}, most of which have originally been proposed by Pfitzmann and Hansen~\cite{pfitzmann2010terminology}: Unlinkability, anonymity, plausible deniability, undetectability, confidentiality, awareness, and policy compliance.

Every PET targets one or more of these goals. Often, it is the case that several goals are targeted by one PET, since they partly overlap or imply each other. For example, it is hardly possible to achieve undetectability without anonymity. In our classification, however, we usually only present one privacy protection goal which we consider to be primarily targeted. % TODO example: k-anonymity, for example, primarily targets unlinkability as a goal, although it is sometimes also attributed to anonymity.

\subsubsection{Privacy Metric}
A technology's suitability and effectiveness need to be measurable to evaluate its added privacy gain and monitor the system over time. 
% We therefore also include metrics that may be used to measure the privacy gain a PET provides. 
Note that the broader problem of creating a metric suite that comprehensively covers the notion of privacy is a research problem out of scope for this paper (see for example Wagner and Yevseyeva~\cite{wagner2021designing}). 
A number of privacy metrics are reviewed in~\cite{wagner2018technical}.

\subsubsection{Functional Scenario}
% Justification comes later when we see that many PETs can be filtered using the functional scenario: Later we show how the functional scenario achieves a better filter than other approaches.
% PETs can be applied in different functional scenarios. For example, homomorphic encryption is a PET applicable to processing scenarios, but which is not meaningful to store and retrieve data. 
The functional context the PET shall be applied in is highly application-specific. Still, some categories of functional scenarios can be identified and can be used as a selection filter. We identify the following generic interactions:\textit{Computation}, \textit{Messaging}, \textit{Retrieval}, \textit{Release}, \textit{Authentication}, as well as \textit{Authorization}.
We use the motivating scenario above to clarify these values: a computation means that data is processed, e.g. a virtual machine processes user data to create recommendations for users; Release is the release of data to another party, e.g. a user sends location data to the social network; Messaging, is a point-to-point interaction between two users via $n$ other parties, e.g. two users of the same social network communicate with each other via the service; Authentication is the process of determining the identity of a user, while Authorization is the process of determining the rights of an identified user.

% \item To that end, we analyze grey literature about popular cloud services to derive representative functional scenarios from them.
% We divide the example service above along its interactions to derive functional scenarios that are applicable in other services as well
% We take apart the interactions of the example service described above to derive generic interaction types that can be found in other services as well.
 % \textit{Storage},
% \item This classification forms the basis for some of the other criteria we propose
% TODO: Release v messaging; messaging involves contextual data
% TODO Note that we do not use a storage scenario, because ... Also, we indicate location-specific PETs

% \subsection{Criterion: Cost}
% % Justification: Spiekermann? Studies for s&p costs?
% \begin{itemize}
% 	\item Two kinds of cost can be considered: set-up and maintenance
% 	\item Values: low (ready-to-use solution), medium (adaptation necessary), high (own implementation necessary)
% \end{itemize}

% \subsection{Criterion: Maintenance}
% \begin{itemize}
% 	\item Values: infrequent (some maintenance is always required), frequent (needs to be re-configured regularly)
% \end{itemize}

\subsubsection{Maturity}
% We would argue that maturity is an indicator for set-up (and maintenance) costs
When selecting a PET, cost factors play an important role, e.g. in the set-up of the PET and in its continuous maintenance. In this paper we use the technology's maturity as an indicator for set-up and maintenance costs, since a technology that is less mature will likely have more defects and will likely imply a more laborious set-up.

For this criterion we loosely base the possible values on the Technology Readiness Level (TRL) which describes a technology's maturity on a scale from 1 to 9~\cite{trl}.
% The concept has first been defined by NASA, but it also being used, e.g., by the European Commission https://ec.europa.eu/research/participants/data/ref/h2020/wp/2014_2015/annexes/h2020-wp1415-annex-g-trl_en.pdf
% \item In this paper we propose a simplified version of the TRL that defines three levels for approaches and implementations: 1 concept stage, possibly with existing PoC; 2 implementation exists, but needs to be adapted; 3 ready-to use, validated solution exists.
We generalize this scale to 3 levels as follows: level 1 is a level often achieved in scientific work, which describes a concept and may already prove feasibility in a proof of concept; level 2 can be seen as the development and testing stage, i.e. adopting such a technology still would require considerable development effort if it is applied to a specific use case; finally, level 3 means that the technology is readily available and field-tested, but may still require some set-up cost for the adaption.

% \subsection{Engagement of Trusted Third Party}
% The necessity to let a PET being operated by a third party influences the trust model of the system and can therefore be a relevant selection criterion. Furthermore, it may incur additional costs. The possible values for this criterion are \textit{required} and \textit{not required}. For example, using a proxy to create an independent access point to a service requires a third party. % TODO better example

% TODO: 1) describe exactly which scenario has which performance requirement (at least high-low) 2) mention that future work needs to propose common performance test guidelines for these scenarios so the performance impact can be measured exactly
\subsubsection{Performance Impact}
The performance of processes and interactions can be impacted by the use of PETs. Evaluating the performance of a certain PET, however, is not trivial, especially in comparison to other technologies. We therefore evaluate a PET's performance in a simplified manner as follows. We first generically describe the performance requirements in a certain functional scenario, and then assess if the use of a PET is expected to significantly impact the performance requirements or not. 

% TODO at least specify magnitude of order?
We consider Computation and Retrieval scenarios generally to have high performance requirements: In these scenarios the user waits for the result of the interaction and will probably notice also small delays. 
In contrast, Authentication, Authorization, and Release scenarios generally have low performance requirements since they are usually one-time actions for which performance impacts are more acceptable.
Also, we consider Messaging to be a scenario of asynchronous communication where small increases in latency are not noticed by the users.

% First, we simply distinguish if a PET impacts the performance or not. Second, we base this evaluation its functional scenario: In retrieval scenarios, for instance, performance can be greatly impacted by small overheads, while in release scenarios this is normally not the case.

\subsubsection{Architectural Impact}
An impact on the architecture is given if the PET requires a dedicated architectural component or modifications to the architecture, e.g. setting up a mix net requires a separate mix server. This is an important selection criterion, since the selection of a PET with this property needs to be considered early on in the design process.

\subsubsection{Utility} 
A utility impact is given if a PET reduces the quality of transactional data, e.g. by distorting or filtering it---and thus decreasing the data's utility.

\rowcolors{2}{SkyBlue!30!gray!60}{white}
\begin{table}
  \caption{Classification of Privacy Enhancing Technologies. A black square indicates that the PET addresses the respective goal, while a triangle indicates a negative impact on the respective criterion.} % It includes a mapping to the privacy protection goals based on LINDDUN
  \centering
  \label{tab:pets}
  \resizebox{\textwidth}{!}{% >{\centering\arraybackslash}m{3cm}
  \begin{tabular}{ >{\centering\arraybackslash}m{3.5cm} | c | c | c | c | c | c | c | >{\centering\arraybackslash}m{2.7cm} | c | c | c | c | c} 
  % \rowcolor{SkyBlue!30!gray!60}
    \toprule
    \multicolumn{1}{c|}{\textbf{\rotatebox[origin=l]{90}{\itshape{Name}}}}
        & \textbf{\rotatebox[origin=l]{90}{\itshape{\makecell{Linkability}}}} 
            & \textbf{\rotatebox[origin=l]{90}{\itshape{\makecell{Identifiability}}}}
                & \textbf{\rotatebox[origin=l]{90}{\itshape{\makecell{Non-Repudiation}}}}
                    & \textbf{\rotatebox[origin=l]{90}{\itshape{\makecell{Detectability}}}}
                        & \textbf{\rotatebox[origin=l]{90}{\itshape{\makecell{Disclosure}}}} 
                            & \textbf{\rotatebox[origin=l]{90}{\itshape{\makecell{Unawareness}}}} 
                                & \textbf{\rotatebox[origin=l]{90}{\itshape{\makecell{Non-Compliance}}}} 
            & \multicolumn{1}{c|}{\textbf{\rotatebox[origin=l]{90}{\itshape{Metrics}}}}
                & \textbf{\rotatebox[origin=l]{90}{\itshape{\makecell{Functional \\Scenario}}}}
                    & \textbf{\rotatebox[origin=l]{90}{\itshape{Maturity}}} 
                        & \textbf{\rotatebox[origin=l]{90}{\itshape{Performance}}} 
                            & \textbf{\rotatebox[origin=l]{90}{\itshape{Architecture}}} 
                                & \textbf{\rotatebox[origin=l]{90}{\itshape{Utility}}}\\ %& \textbf{Maintenance} & \textbf{TTP} \\
    \midrule
    k-anonymity, l-diversity, t-closeness
                            & $\blacksquare$ &   &   &   &   &   &   & Data similarity & Release & 3~\cite{arx} & & & $\blacktriangleleft$\\ % it is primarily targeted at unlinkability via equivalence classes; identifiability should be targeted differently, e.g. via filters; maturity shown e.g. by https://cloud.google.com/dlp/docs/compute-k-anonymity
    % Mix Zone                & $\blacksquare$ &   &   &   &   &   &   & Data similarity & Release (Location) &              & & $\blacktriangleleft$ & $\blacktriangleleft$ \\ % Its main purpose is to unlink location  data points, so anonymity is not primarily targeted; it can be seen as a location-specific application of the k-anonymity concept
    Suppression             & $\blacksquare$ & $\blacksquare$ &   &   &   &   &   & Data similarity & Release & 3~\cite{arx}   & & & $\blacktriangleleft$ \\ % TODO: recoding, suppression, aggregation, swapping, noise masking, PRAM, synthetic data % \footnote{We do not provide a reference here, since filters are simple functions that do not require a dedicated tool.}
    Recoding                & $\blacksquare$ & &   &   &   &   &   & Data similarity & Release & 3~\cite{arx} 
                                                                                                      & & & $\blacktriangleleft$ \\
    Aggregation             & $\blacksquare$ & &   &   &   &   &   & Data similarity & Release & 3~\cite{arx}                                                                                                  & & $\blacktriangleleft$ & $\blacktriangleleft$ \\ % TODO redundant?
    Swapping                & $\blacksquare$ & &   &   &   &   &   & Data similarity & Release & 1~\cite{hundepool2012statistical} 
                                                                                                      & & & \\ 
    Noise masking           & $\blacksquare$ & &   &   &   &   &   & Data similarity & Release & 3~\cite{mivule2013utilizing}                                                                              & & & $\blacktriangleleft$ \\ 
    PRAM                    & $\blacksquare$ & &   &   &   &   &   & Data similarity & Release & 2~\cite{pram}& & & $\blacktriangleleft$ \\ 
    Synthetic data          & $\blacksquare$ & &   &   &   &   &   & Data similarity & Release & & & & $\blacktriangleleft$ \\ 
    % Functional encryption
    Mix Network             & $\blacksquare$ & $\blacksquare$ &   & $\blacksquare$ &   &   &   &                 & Messaging & 3~\cite{mix-net-tor} & $\blacktriangleleft$ & $\blacktriangleleft$ & \\ % A mix net can disguise the identity of the sender; as such it can also provide unlinkability, e.g. two requests cannot easily be linked to the same person; similarly, provides deniability, but this is secondary
    Group Signatures        & $\blacksquare$ & $\blacksquare$ & $\blacksquare$ &   &   &   &   & Cryptographic Games & Release  & 2~\cite{group-signatures} & & $\blacktriangleleft$ & \\ % preserve anonymity, but allow signing; practically unlinks several signatures from the same group; allows to deny having signed a message (though it is on behalf of the group)
    Anonymous Credentials   & $\blacksquare$ & $\blacksquare$ &   &   &   &   &   & Cryptographic Games & AuthN       & 2~\cite{anonymous-credentials} & & & \\ % = self-sovereign identities?
    Zero Knowledge Proofs   & $\blacksquare$ & $\blacksquare$ &   &   &   &   &   & Cryptographic Games & AuthN, AuthZ& 2~\cite{zkp} & $\blacktriangleleft$ & & \\ % tbd: +revocation (e.g. blacklists)
    Pseudonymization        &   & $\blacksquare$ &   &   &   &   &   & Entropy & \makecell{AuthN, \\Release} & 3~\cite{european2019pseudonymisation} & & & \\ % TODO ; covers LINDDUN's "Privacy-enhancing identity management system"
    % Spatial Obfuscation/Cloaking 
                            % & $\blacksquare$ &   &   &   &   &   &   & Data similarity & Release (Location)    & 1 & & & \\ % TODO architecture component?
    Deniable Authentication & & & $\blacksquare$ &   &   &   &   & Cryptographic Games & Messaging & 3~\cite{otr} & & & \\ % the PETs that target N are actually preserving it rather than introducing it
    Deniable encryption     &   &   & $\blacksquare$ &   & $\blacksquare$ &   &   & Cryptographic Games & Messaging  & 3~\cite{otr} & & & \\
    Searchable Encryption   &   &   &   &   & $\blacksquare$ &   &   & Cryptographic Games & Retrieval  & 3~\cite{opensse} & & & \\ % Storage?
    Private Information Retrieval
                            &   &   &   &   & $\blacksquare$ &   &   & Cryptographic Games & Retrieval  & 2~\cite{muchpir} & $\blacktriangleleft$ & $\blacktriangleleft$ & \\ % confidentiality of the transactional data, i.e. the requested data; may also target linkability of transactional data
    Oblivious Transfer      &   &   &   &   & $\blacksquare$ &   &   & Cryptographic Games & Retrieval  & 2~\cite{12ot}    & $\blacktriangleleft$ & $\blacktriangleleft$ & \\ % see above
    Proxy Re-Encryption     &   &   &   &   & $\blacksquare$ &   &   & Cryptographic Games & Messaging  & 2~\cite{proxyre} & $\blacktriangleleft$ & $\blacktriangleleft$ & \\ % This PET preserves the confidentiality of encrypted data while altering it; as such it does not directly "create" confidentiality but preserves it considering in the scenario where encrypted data need to be transformed
    Homomorphic Encryption  &   &   &   &   & $\blacksquare$ &   &   & Cryptographic Games & Computation & 2~\cite{googlehe} & $\blacktriangleleft$ & & \\ % key length? % ~\cite{googlehe,seal} https://zama.ai/ https://palisade-crypto.org/ https://kumu.io/iliailia/fhe-graph % \footnote{The functionality that homomorphic encryption offers currently is limited to a few simple operations.}
    Trusted Execution Environment 
                            &   &   &   &   & $\blacksquare$ &   &   & Cryptographic Games & Computation & 3~\cite{intelsgx,amdsev} & & & \\
    (A)Symmetric Encryption &   &   &   &   & $\blacksquare$ &   &   & Cryptographic Games & \makecell{Messaging,  \\Release}  & 3 & & & \\
    Dummy traffic           &   &   &   & $\blacksquare$ &   &   &   & Data similarity     & Messaging & & & & \\ % metric is binary if the attacker cannot distinguish between messages; but minimum of messages is required...
    % Private Messaging       &   &   & $\blacksquare$ &   & $\blacksquare$ &   &   & Cryptographic Games & Messaging & 3~\cite{otr} & & & \\
    Steganography           &   &   &   & $\blacksquare$ & $\blacksquare$ &   &   & Entropy & Messaging   & 2~\cite{steganography}  & & & \\ % TODO
    MPC                     &   &   &   &   & $\blacksquare$ &   &   & Cryptographic Games & Computation & 3~\cite{mpc} & $\blacktriangleleft$ & & \\ % performance impact due to circuit creation (similar reasoning as ZKP) % "here, a limitation of the unlinkability goal becomes apparent: unlinking a person from a function input is different than unlinking two actions or data items"
    % DC-networks vs MPC (secure function evaluation)?
    % Privacy-Preserving Matching
    %                         &   &   &   &   & $\blacksquare$ &   &   & Cryptographic Games & Computation & & & & \\ % may also target LI of respective persons; TODO for implementations also check ``linkage'' etc.: https://gsm.ucdavis.edu/sites/default/files/2020-10/design_and_implementation_of_a_privacy_preserving_electronic_health_record_linkage_tool_in_chicago.pdf
    Local Differential Privacy    &  &   & $\blacksquare$ &   &   &   &   &Indistinguishability& Release  & 2~\cite{diffpriv} & & & $\blacktriangleleft$\\ % TODO adjust library
    % Local Differential Privacy    &  &   & $\blacksquare$ &   &   &   &   &Indistinguishability& Release (Location)    &  & & & $\blacktriangleleft$\\ % TODO library
    Global Differential Privacy   & $\blacksquare$ &   &  &   &   &   &   &Indistinguishability& Release  & 2~\cite{diffpriv} & & & $\blacktriangleleft$\\ % unlinkability of related datasets % TODO is this Release or retrieval? Release means protect the people in the dataset; retrieval means protect the user retrieving sth
    % RBAC, ABAC            &   &   &   &   & X &   & X & & & \\ % Policy enforcement: usage control, access control (RBAC, ABAC, zt), authN, authZ
    % Usage Control         &   &   &   &   & X &   & X & & & \\
    % Commitment scheme     &   &   &   &   & X &   &   & \textit{binary} & Release  & \\ % rather practical consideration
    % Covert channels https://dl.acm.org/doi/10.1145/1315245.1315283
    % Attribute-based creds.  & $\blacksquare$ & $\blacksquare$ &   &   & $\blacksquare$ &   & $\blacksquare$ & Cryptographic Games & AuthN, AuthZ& 2~\cite{abc} & & & \\ % covers LINDDUN's "User-controlled identity management system"
    Attribute-based encr.   &   &   &   &   & $\blacksquare$ &   & & Cryptographic Games & AuthN, AuthZ& 2~\cite{abe} & & & \\
    % CP Attribute-based encryption & X & X &   &   &   &   &   & \textit{binary}& AuthN, AuthZ& \\
    % KP Attribute-based encryption & X & X &   &   &   &   &   & \textit{binary}& AuthN, AuthZ& \\
    % Anonymous buyer seller watermark protocol https://www.esat.kuleuven.be/cosic/publications/article-1235.pdf; this is basically homomorphic encryption (or MPC...) --> Georg: is this HE or does this even have to do something with privacy? Seems more like non-repudiation or verifiability...
    Federated Learning      &   &   &   &   & $\blacksquare$ &   &   & Attacker Success Probability & Release & 2~\cite{ftt} & & $\blacktriangleleft$ & \\
    \bottomrule
  \end{tabular}}
\end{table}

\subsection{Classifying PETs}
Table~\ref{tab:pets} shows our classification proposal according to the criteria defined above. Note that our classification only includes the so-called \textit{hard privacy goals}~\cite{deng2011privacy}, i.e. Anonymity, Unlinkability, Plausible Deniability, and Undetectability. The \textit{soft privacy goals} Awareness and Policy Compliance\footnote{As \textit{soft privacy goals}, some works also use the goals Intervenability and Transparency~\cite{hansen2015protection}.} are usually targeted by more generic design patterns (see e.g.~\cite{al2021land}). There can, however, be overlaps between technologies and patterns: For instance, onion routing can be seen both as a design pattern and a PET.

Note that regarding the targeted privacy protection goal, we always assign the goal that is targeted \textit{primarily}. For example, Zero Knowledge Proofs (ZKP) primarily address the threats linkability and identifiability. While they could theoretically also be used to secretly release information, an engineer would not choose ZKP to achieve confidentiality.

Note also that throughout the paper, we use \textit{k-anonymity}~\cite{sweeney2002k} as a placeholder also for other related ones like l-diversity~\cite{machanavajjhala2007diversity}, t-closeness~\cite{li2007t}, etc.
% For example, private information retrieval (PIR) primarily targets disclosure of information. While it also implies that requests using a PIR protocol are unlinkable, an engineer would not choose PIR to achieve unlinkability.
% The cryptographic games metrics~\cite{wagner2018technical} refer to game theoretical challenge-response interactions.

\section{Use Case and Discussion}
\label{sec:discussion}
%!TEX root = ../selecting-pets.tex

\subsection{Use Case}
In this section, we compare our classification to the LINDDUN classification which was first proposed by Deng et al.~\cite{deng2011privacy} and has since been updated on the LINDDUN website~\cite{linddunwebsite}\footnote{Note that we do not compare our approach to Heurix et al.~\cite{heurix2015taxonomy}, since they partly use different privacy protection goals and generally provide few selection criteria.}.
% \begin{itemize}
% 	\item To compare the two we align the set of input PETs, apply the two classifications to an example service, and then discuss the results regarding quantity and quality aspects. 
% \end{itemize}
Note that the LINDDUN classification also includes a selection methodology. This methodology is based on the LINDDUN threat types which are connected to general mitigation strategies. These strategies in turn are mapped to applicable PETs. For instance, a linkability threat concerning a data flow may be mapped to the mitigation strategy \textit{protect transactional data} which in turn yields the PETs multi-party computation, encryption, and others.

We use again the motivating example introduced earlier, and extend and detail it as follows: our example cloud service is a social network that allows to add friends, exchange private messages with each other, as well as make public posts. Furthermore, the service offers a location-based feature where users can provide their location data and are then offered possible contacts in their proximity they can message.
Note that the example used in the original LINDDUN approach is a social network application as well, making it a well-suited basis for a comparison.

% TODO figure(s): DFD

% \begin{itemize}
% 	To make the two methodologies comparable, we align the set of PETs that is used as the input for the selection. Since our classification contains more PETs than LINDDUN, we add two of them to the LINDDUN classification, especially since it does not account for location-specific PETs, and we omit remaining ones. We add PETs as follows: % could be part of appendix
% 	\textit{Spatial obfuscation / cloaking}, as well as \textit{position dummies}, are techniques that can be mapped to the LINDDUN mitigation strategy \textit{concealing association}, its branch \textit{protect transactional data}, and the category \textit{generalize}.
% \end{itemize}

% \item To make the two methodologies comparable, we align the set of PETs that is used as the input for the selection. Since our classification contains more PETs than LINDDUN, we add some of them to the LINDDUN classification, and we omit remaining ones. We add PETs as follows: % TODO 

We use three example threats from different LINDDUN categories, i.e. a linkability, an identifiability, and a disclosure threat, to demonstrate the effectiveness of our classification in comparison to the LINDDUN classification. Note that these threats have also been identified (on a more high level) in an example analysis conducted by the LINDDUN authors for their social network running example, see~\cite{Wuyts}. 
The threats and results from the PET-selection of both approaches are described in the following. Table~\ref{tab:comparison} summarizes the results.

% TODO: align with the table after re-checking everything
\subsubsection{Identifiability}
This threat describes the possibility that the server can identify the user via the transmitted transactional data, e.g. due to identifiers like name or address.

\begin{itemize}
	% \item This threat concerns the loss of anonymity of the user against an outside attacker, e.g. due to eavesdropping on the connection between user and server. % This is rather Release
	% \item This threat describes the possibility that the service provider can identify the user via the transmitted metadata, like IP address, etc. % This is rather linkability
	\item Applying LINDDUN in this scenario we obtain 7 PETs. With our classification we obtain: Suppression, Recoding, Aggregation, Swapping, Noise Masking, PRAM, Synthetic Data, Mix Network, Group Signatures, Global Differential Privacy.
	\item This is the only case in which our classification outputs more PETs than the LINDDUN methodology.
\end{itemize}

\subsubsection{Linkability}
This threat concerns potential linkability of different types of transactional data which is released by the user to the server. Applying the LINDDUN method results in a set of 10 PETs as summarized in Table~\ref{tab:comparison}.
% \textit{Privacy enhancing identity management system}, \textit{user-controlled identity management system}, \textit{privacy-preserving biometrics}, \textit{privacy authentication}, \textit{anonymous credentials}, MPC, watermarking protocol, \textit{(A)symmetric encryption}, \textit{deniable encryption}, \textit{homomorphic encryption}, \textit{verifiable encryption}, \textit{k-anonymity} and \textit{l-diversity}.

In comparison we can see that our classification correctly omits several PETs since they are not usable to mitigate a linkability threat, i.e. Multi-Party Computation, (A)Symmetric Encryption, and Homomorphic Encryption. 
Furthermore, it does provide more applicable ones, e.g. Aggregation and Noise Masking which can obfuscate the relation between data. % TODO more citations
%k-anonymity and l-diversity, Suppression, Recoding, Aggregation, Swapping, Noise Masking, PRAM, Synthetic Data, Group Signatures, and Global Differential privacy.

% Our classification, in contrast, results in only two PETs: \textit{position dummies} and \textit{spatial obfuscation/cloaking}, i.e. either a misinformation or a generalization technique. Since we do not classify encryption techniques as unlinkability techniques, they are not suggested for mitigating this threat: encryption only protects unlinkability of transactional data, and only as long as it stays encrypted---which is not a realistic scenario in a common user-server interaction.

\subsubsection{Disclosure}	
This threat concerns the disclosure of transmitted data due to insecure connections. Applying the LINDDUN method results in a set of 12 PETs, see Table~\ref{tab:comparison}. 
% \textit{(A)symmetric encryption}, \textit{deniable encryption}, \textit{homomorphic encryption}, \textit{verifiable encryption}, \textit{context-based access control}, \textit{privacy-aware access control}, \textit{private information retrieval}, \textit{oblivious transfer}, \textit{privacy-preserving data mining}, \textit{searchable encryption}, \textit{private search}, \textit{k-anonymity} etc.
Our approach results in 4 PETs: Proxy Re-Encryption, Deniable Encryption, (A)symmetric Encryption, and Steganography. % again the functional scenario is helpful: disclosure implies that the data \textbf{needs} to be shared and cannot simply be encrypted or protected via access control

% \subsubsection{Detectability}
% % TODO: (1) integrate LINDDUN PETs in our classification (2) adapt LINDDUN threats (e.g. DC nets are MPC)
% A \textit{detectability} threat occurs on messaging data flows when users exchange messages with each other, i.e. the service provider can observe that messages are being exchanged, which can reveal potentially sensitive contacts. Applying the (adapted) LINDDUN classification and methodology, this threat can be seen as contextual\footnote{On the LINDDUN website, the authors state that detectability threats are ``entirely focused on contextual data''~\cite{linddunwebsite}.}. We then obtain as possible PETs \textit{Dummy traffic}, \textit{MPC}, \textit{Steganography}, \textit{Dummy traffic}, \textit{Covert communication}\footnote{Note that LINDDUN references the covert communication PET with~\cite{moskowitz2003covert}, where this is discussed in the context of mix networks.}, and \textit{Spread spectrum}.
% In contrast, our classification only results in dummy traffic, steganography, and mix network---other PETs are omitted due to the functional scenario.

% \item Note that without restricting the input PETs to the ones also classified in LINDDUN, the results are similar.

% TODO claim
\begin{itemize}
	% Quality / Quantity
 	\item On the basis of this comparison, we expect that our classification better supports engineers in the selection of PETs than existing classifications and selection approaches. % we would argue that our classification produces more precise results, mainly due to the contextualization of PETs using their functional scenario. 
 	\item This is achieved mainly by filtering through the functional scenario, but also the more targeted mapping of PETs to protection goals.
 	% Also, it does not omit any usable PETs. 
 	% Also, the PETs it results in are more 
 	% TODO align "last example"
 	\item Consider also that using our classification, a user can further prioritize the results using the maturity, utility, and architecture impact criteria. For example, the (a)symmetric encryption may be prioritized in the last example because it has high maturity and no impact on architecture or utility.
\end{itemize} 

\rowcolors{2}{SkyBlue!30!gray!60}{white}
% \rowcolors{2}{gray!25}{white}
\begin{table}
	\caption{Comparison of the results of applying the LINDDUN PET Selection method and our classification. Those PETs that are included in both classifications are written in bold. For example, for the linkability threat LINDDUN suggests Multi-Party Computation as a possible mitigation. This PET is also included in our classification, which, however, has not suggested it for this threat. Verifiable Encryption is also suggested by LINDDUN, but it has not been considered in our classification at all.}
	\centering
	\label{tab:comparison}
	\resizebox{\textwidth}{!}{%
	\begin{scriptsize}% or footnotesize, scriptsize, tiny, etc.
	\begin{tabular}{p{0.14\linewidth} | p{0.45\linewidth} | p{0.35\linewidth}}
  % \rowcolor{gray!25}{}
  	\toprule
    \textbf{Threat} & \textbf{LINDDUN Result} & \textbf{Our Classification} \\
    \hline
    % Detectability 	& Dummy traffic, Multi-Party Computation, Steganography, Dummy traffic, Covert communication, Spread spectrum 
    % 										& Dummy traffic, Steganography, Mix network \\
    % Linkability 		& \makecell[l]{Privacy Enhancing Identity Management System \\ User-Controlled Identity Management System \\ Privacy-Preserving Biometrics \\ Privacy Authentication, Anonymous Credentials \\ Multi-Party Computation \\ Watermarking Protocol \\ (A)symmetric encryption \\ Deniable Encryption \\ Homomorphic Encryption \\ Verifiable Encryption \\ k-anonymity and l-diversity}
    % 											& \makecell[l]{Position Dummies \\ Spatial Obfuscation/Cloaking}\\
    % Deprecated: \makecell[bl]{Linkability\\\\\\\\\\\\} 		& \makecell[bl]{\textbf{k-anonymity and l-diversity} \\ Privacy-Enhancing Identity Managem. Syst. \\ User-Controlled Identity Managem. Syst. \\ Privacy-Preserving Biometrics \\ Privacy Authentication \\ Anonymous Credentials \\ Multi-Party Computation \\ Watermarking Protocol \\ (A)symmetric encryption \\ Deniable Encryption \\ Homomorphic Encryption \\ Verifiable Encryption}
    \makecell[bl]{Linkability\\\\\\\\\\\\} 		& \makecell[bl]{\textbf{k-anonymity} \\ \textbf{Multi-Party Computation} \\ \textbf{(A)symmetric encryption} \\ \textbf{Homomorphic Encryption} \\ \textbf{Deniable Encryption} \\ Anonymous Buyerseller Watermarking Prot. \\ Verifiable Encryption \\\\\\\\}
    											& \makecell[bl]{\textbf{k-anonymity and l-diversity} \\ Suppression \\ Recoding \\ Aggregation \\ Swapping \\ Noise Masking \\ PRAM \\ Synthetic Data \\ Group Signatures \\ Global Differential privacy} \\
    \hline
    Disclosure (Release) 	
    								& \makecell[l]{\textbf{k-Anonymity} \\ \textbf{(A)symmetric Encryption} \\ \textbf{Homomorphic Encryption} \\ \textbf{Private Information Retrieval} \\ \textbf{Oblivious Transfer} \\ \textbf{Searchable Encryption} \\ \textbf{Deniable Encryption} \\ Verifiable Encryption \\ Context-Based Access Control \\ Privacy-Aware Access Control \\ Privacy-Preserving Data Mining \\ Private Search}
    											& \makecell[l]{\textbf{(A)Symmetric Encryption} \\ Federated Learning \\\\\\\\\\\\\\\\\\\\\\} \\
    											% & \makecell[l]{\textbf{k-anonymity and l-diversity} \\ \textbf{(A)symmetric Encryption} \\ Proxy Re-Encryption \\ Steganography\\\\\\\\\\\\\\\\} \\
    \hline
    % Computation
    Disclosure (Computation)
    								& \makecell[l]{\textbf{Homomorphic Encryption} \\ \textbf{k-Anonymity} \\ \textbf{(A)symmetric Encryption} \\ \textbf{Private Information Retrieval} \\ \textbf{Oblivious Transfer} \\ \textbf{Searchable Encryption} \\ \textbf{Deniable Encryption} \\ Verifiable Encryption \\ Context-Based Access Control \\ Privacy-Aware Access Control \\ Privacy-Preserving Data Mining \\ Private Search}
    											& \makecell[l]{\textbf{Homomorphic encryption} \\ \textbf{Multi-Party Computation} \\ Trusted Execution Environment \\\\\\\\\\\\\\\\\\\\} \\
		\hline
    Identifiability (Release) & \makecell[l]{\textbf{Multi-Party Computation} \\ \textbf{(A)symmetric encryption} \\ \textbf{Homomorphic encryption} \\ \textbf{Deniable Encryption} \\ Anonymous Buyerseller Watermarking Prot. \\ Verifiable Encryption\\}
         									& \makecell[l]{\textbf{k-anonymity and l-diversity} \\ Suppression \\ Pseudonymization \\ Group Signatures \\ Global Differential Privacy\\\\} \\
		% Identifiability (Messaging) 
		% 								& \makecell[l]{\textbf{Multi-Party Computation} \\ \textbf{(A)symmetric encryption} \\ Anonymous buyerseller watermarking prot. \\ Deniable encryption \\ Homomorphic encryption \\ Verifiable Encryption\\\\\\\\\\}
  %        									& \makecell[l]{\textbf{Mix Network}} \\
    % TODO: missing scenario messaging
    \bottomrule
	\end{tabular}
	\end{scriptsize}}
\end{table}

\subsection{Discussion}
\subsubsection{Limitations}
% Comprehensiveness
One limitation of our approach is its coverage: while it is an extensive collection of 32 PETs, it is not complete and should be extended and maintained in the future. 
% Timeliness
Especially the evaluation of the maturity criterion may become outdated soon, e.g. if current research proposals are developed further.
% TODO: We expect, however, that most evaluations can serve as a helpful indicator for several years to come

% TODO: can LINDDUN threat trees be mapped to our table?
% Criteria
Furthermore, the criteria we propose are deduced from a use case. Therefore, their effectiveness still has to be validated in real-world studies. Some relevant criteria, e.g. regarding other non-functional requirements, could also be missing. 
% This derivation does not guarantee that they are actually useful in a real-world application scenario.

Still, we expect our classification to improve the systematic selection of PETs, and the evaluation of software architectures. 
For instance, design decisions in software architectures can be linked to our classifications and systematically evaluated. % TODO how?
% For example, requirements could be mapped to the classification to document how a certain requirement has been addressed.

Our classification could be biased since the use case and the threats the LINDDUN analysis identifies were known to the authors before the classification was finished. We assume, however, that the bias is low since it was developed in discussion with multiple domain experts who did not know the LINDDUN analysis. Also, we would argue that it is evident from that both the general criteria as well as the classification itself are independently applicable from the social network example.%\footnote{Note, however, that the authors did not know about the LINDDUN example until the classification was finished.}

% \begin{itemize}
	% \item Many PETs can only meaningfully be measured by cryptographic games, which is usually not too useful for a practical user
	% \item Also, the improved precision potentially allows to use it in a (semi-)automated support system.
	% \item Furthermore, it can be used in threat modeling tools to suggest mitigative PETs.
% \end{itemize}

\subsubsection{Criteria}
Evidently, it is not guaranteed that the criteria we propose are comprehensive and that they capture what engineers require as selection criteria in practice. On the basis of the case example above, however, we expect that it works better than existing approaches also in other applications.

In comparison to Al-Momani et al.~\cite{al2021land}, we do not include criteria that indicate an impact on security and complexity, because we would argue that they are redundant. Complexity is always increased by a PET to some degree, while the actual degree of complexity is too difficult to measure.
With regards to the security impact, there is one privacy protection goal that directly contradicts a security goal, i.e. plausible deniability contradicts non-repudiation. Thus, any PET that targets plausible deniability also counteracts said security goal which can therefore directly be derived from our classification.

% \item The effectiveness of PETs targeting disclosure cannot easily be measured since disclosure is binary: a datum is either disclosed or hidden. Here, only the hiding strength, e.g. the encryption strength, can be measured.

% TODO: ideally we also compare the selection above to Heurix
In the following we also compare our criteria with Heurix et al.~\cite{heurix2015taxonomy}:
\begin{itemize}
	\item The \textit{Aim} dimension is similar to our privacy protection goal.
	\item The \textit{Scenario} dimension is not in scope for us, since we focus on client-server interactions where the server is untrusted. % TODO: although there are exceptions when it comes to disclosure
	\item The \textit{Aspect} dimension is similar to the mitigation strategies in the LINDDUN method~\cite{linddunwebsite}.
	We do not consider these because they are implied in the functional scenario: for instance, a PET that targets the authentication scenario addresses protection of ID. 
	\item The \textit{Foundation} and \textit{Data} dimensions not relevant for the selection in practice.
	\item The \textit{Trusted Third Party} dimension is covered by our criterion of architectural impact.
	\item The \textit{Reversibility} dimension is largely the same criterion as our utility criterion, since the distortion or deletion of a value is usually non-reversible.
\end{itemize}

Furthermore, previous selection methodologies do not provide means for prioritizing PETs~\cite{kunz2020selecting,deng2011privacy}. In a set of potentially applicable PETs, however, we would argue it is important to have prioritization factors, such as their maturity, as we propose in this paper.

\section{Related Work}
\label{relatedwork}
%!TEX root = ../selecting-pets.tex

\subsection{Privacy By Design}
Generally, our classification can be seen as a tool that supports privacy by design. As such, it is complementary to other privacy engineering methods which often assume a PET-selection without further detailing this step~\cite{gurses2011engineering,gurses2015engineering,spiekermann2008engineering}.
One such approach is proposed by Alshammari and Simpson~\cite{alshammari2018privacy} who develop an engineering process that devises architectural strategies, i.e. combinations of tactics, patterns, and PETs, to fulfill privacy goals. In their approach, the set of usable PETs is determined by the chosen design pattern. The concrete selection of a PET, however, is not specified in their work. As such, our classification could be integrated into their methodology.
	
% PRIPARE, Kung~\cite{kung2014pears} % Antonio Kung. PEARs: Privacy Enhancing ARchitectures. In Privacy Technologies and Policy – Second Annual Privacy Forum, APF 2014, Athens, Greece, May 20-21, 2014. Proceedings, pages 18–29, 2014

\subsection{Systematization of PETs}
% \item LINDDUN threat modeling \cite{deng2011privacy,wuyts2020linddun}
Also, further works have investigated the selection and systematization of PETs. 
Al-Momani et al.~\cite{al2021land} follow a similar approach as we do but focus on privacy patterns rather than concrete technologies. As explained above, patterns rather target soft privacy goals. They use the following criteria to classify patterns: applicability scope %(architecture- or implementation-specific)
, privacy objective, qualities (e.g. performance impact), data focus, and LINDDUN GO hotspot. In this paper, we have partly used similar criteria; many criteria, however, are different since the selection of concrete technologies requires other seleciton criteria than patterns, e.g. maturity. Note also, that our classification of the targeted privacy objective (called privacy goal in this paper) differs in some cases from theirs.
% One of their criteria, for instance, is \textit{complexity} which is evaluated with \textit{yes} or \textit{no}. Yet, every PET increases complexity to some degree.
% Note that most of their patterns target what LINDDUN calls \textit{soft privacy} goals, e.g. transparency. Generally, patterns rather target these soft privacy goals, while technologies rather target hard privacy goals. % There are still overlaps, which, however, may not be problematic.
Kunz et al.~\cite{kunz2020selecting} also propose a selection method for PETs but their approach is limited to PETs that manipulate transactional data, e.g. generalization or filter.
A systematization of PETs is proposed by Heurix et al.~\cite{heurix2015taxonomy}. They identify the dimensions Scenario, Aspect, Aim, Foundation, Data, trusted third party, and Reversibility. 
% TODO: 
We would argue, however, that these dimensions are not directly helpful for engineers who have to select a concrete PET.
% This classification also overlaps with ours which also includes the TTP criterion.
% \item Baca and Petersen~\cite{baca2010prioritizing}: countermeasures
% TODO https://scholar.google.com/scholar?hl=de&as_sdt=0%2C5&q=selecting+pets+%7Eprivacy&btnG=
Rubio et al.~\cite{rubio2016selecting} review 10 PETs regarding their efficiency for smart grids. Since their analysis is focused on smart grids, they also use respective classification criteria, like suitability for billing or monitoring purposes. Their work is thus complementary to ours since we do not include smart grids as a functional scenario.
% The PETs they review are, however, generic categories of technologies, like trusted computation or perturbation, rather than concrete instances of technologies. 
% in that sense they also analyze them w.r.t. their functional scenario but focus on smart grids only

ENISA has previously promoted a prototype of a PET maturity repository~\cite{european2019petsmaturity}; to the best of the authors' knowledge, however, ENISA has not continued this repository. 

There is furthermore an ENISA publication about Privacy and Data Protection by Design which classifies PETs into several categories which, however, are not intended as selection criteria~\cite{european2019pbd}.
% "Further to these characteristics, additional categorization can be performed with regards to the GDPR data protection principles that each category can support, at least in theory. Attempting to perform such a taxonomy could be of great value to data controllers and processors as it would provide a reference model of either what purposes each tool or technique can serve or as an indication of what is already achieved by already deployed tools and techniques"

% https://www.private-ai.ca/PETs_Decision_Tree.svg
% https://www.ngi.eu/discover-ngi-solutions/

% TODO: how are these connected to ours? utility-disclosure-?
Another recent ENISA publication~\cite{european2022dpe} proposes a categorization of PETs regarding the categories \textit{truth-preserving}, \textit{intelligibility-preserving}, and \textit{operable technology}.

% \begin{itemize}
% 	\item LINDDUN GO~\cite{wuyts2020linddun} also defines \textit{hotspots} which define where in the system a threat occurs, e.g. in a storage or in an in- or outbound data flow. These partly overlap with the functional scenarios we propose. % TODO These also include storages, but they rather hint at the location of the threat than the functional context.
% \end{itemize}

\section{Conclusions}
\label{conclusions}
%!TEX root = ../selecting-pets.tex

The selection of privacy-enhancing technologies is a task that is difficult to address systematically. 
% Previous works have approached this problem from different perspectives~\cite{alshammari2018privacy,heurix2015taxonomy,kunz2020selecting}.
In this paper we have proposed application-oriented criteria that allow such a systematic selection, and have classified a number of PETs according to these criteria, e.g. their functional scenario and applicable metrics.

One open issue is the performance evaluation of PETs, since their performances are usually not easily comparable. In future work, we therefore plan to propose an evaluation framework for the measurement of the performance of PETs.
We also want to extend our classification with more PETs, and connect them with other concepts, such as design patterns. Unifying these in, e.g., a comprehensive ontological description of privacy concepts may represent a valuable support tool for engineers.
% Furthermore, we want to develop a web application that provides better usability of the classification, e.g. in the form of a decision tree. 
Furthermore, existing threat modeling tools can be extended with suggestions for mitigation based on our classification.
Future work also needs to show the effectiveness of the proposed classification in real-world applications.

% Comparing our classification with~\cite{al2021land}, it is striking that technologies rather target LINDD threats (or their corresponding goals), while patterns rather target intervenability and transparency. In the future it could be a valuable contribution to 

% \subsubsection{Acknowledgements} Please place your acknowledgments at
% the end of the paper, preceded by an unnumbered run-in heading (i.e.
% 3rd-level heading).

\bibliographystyle{splncs04}
\bibliography{references.bib}

\end{document}